\documentclass[aip, linenumbers, reprint, longbibliography]{revtex4-2}

\draft 
\usepackage{graphicx}
\usepackage{subfigure}
\usepackage{ulem}
\usepackage{color}
\usepackage{bm}
\usepackage{comment}
\usepackage{soul}
\usepackage{amsmath,amssymb}

\usepackage{todonotes}
\definecolor{electricpurple}{rgb}{0.75, 0.0, 1.0}

\definecolor{green}{RGB}{0,154,0}
\definecolor{purple}{RGB}{128,0,128}
\definecolor{Marco}{rgb}{0.6275, 0.1255, 0.9412}

\usepackage{lineno}

\begin{document}

\nolinenumbers


\title{Hierarchical meta-porous materials as sound absorbers}

\author{S. Kuznetsova}
\affiliation{Univ. Lille, CNRS, Centrale Lille, Junia, Univ. Polytechnique Hauts-de-France, UMR 8520 - IEMN - Institut d’Electronique de Micro\'electronique et de Nanotechnologie, F-59000 Lille, France}
\author{S. Deleplanque}%
\affiliation{Univ. Lille, CNRS, Centrale Lille, Junia, Univ. Polytechnique Hauts-de-France, UMR 8520 - IEMN - Institut d’Electronique de Micro\'electronique et de Nanotechnologie, F-59000 Lille, France}
\author{B. Dubus}%
\affiliation{Univ. Lille, CNRS, Centrale Lille, Junia, Univ. Polytechnique Hauts-de-France, UMR 8520 - IEMN - Institut d’Electronique de Micro\'electronique et de Nanotechnologie, F-59000 Lille, France}
\author{M. Miniaci}%
\email{marco.miniaci@univ-lille.fr}
\affiliation{Univ. Lille, CNRS, Centrale Lille, Junia, Univ. Polytechnique Hauts-de-France, UMR 8520 - IEMN - Institut d’Electronique de Micro\'electronique et de Nanotechnologie, F-59000 Lille, France}

\date{\today}

\begin{abstract}
The absorption of sound has great significance in many scientific and engineering applications, from room acoustics to noise mitigation.
In this context, porous materials have emerged as a viable solution towards high absorption performance and lightweight designs.
However, their performance is somehow limited in the low frequency regime.
\\
Inspired by the concept of recursive patterns over multiple length scales typical of many natural materials, here, we propose a hierarchical organization of multilayered porous media and investigate their performance in terms of sound absorption.
\\
Two types of designs are investigated: a hierarchical periodic and a hierarchical gradient.
In both cases it is found that the introduction of multiple levels of hierarchy allows to simultaneously (i) increase the level of absorption compared to the corresponding bulk block of porous material, along with (ii) a reduction of the quantity of porous material required.
Both the cases of normal and oblique incidences are examined.
\\
The methodological approach is based on the transfer matrix method, optimization algorithms (metaheuristic Greedy Randomized Adaptive Search Procedure), and finite element calculations.
An excellent agreement is found between the analytical and the numerical simulations.
\end{abstract}
\maketitle 

\clearpage
\section{Introduction}
\label{Introduction}
One of the stumbling blocks in modern acoustics is the absorption of the sound at the low frequencies, especially when the thickness of the absorbing structure is smaller than the wavelength of the incident sound wave.
This performance, highly desirable in most of the noise shielding applications related to human activities, is even more chased in contexts such as room acoustics, engine noise control, and, in general, where the space to host the absorber is limited~\cite{bies2017engineering}.
\\
In this context, porous materials have emerged as a viable solution towards high absorption performance and lightweight designs~\cite{cao2018porous}.
The main mechanism responsible for the sound absorption in porous media derives from the conversion of the sound energy into heat, further dissipated through the large number of complex micro-pores of which these materials are composed~\cite{wang2021broadband}.
Several theoretical models to describe their acoustic behavior have been proposed, so far~\cite{biot1956theoryI, biot1956theoryII, allard1986acoustical, johnson1987theory, jimenez2021acoustic}.
Among them, the Johnson-Champoux-Allard (JCA) model is probably the currently most widely used~\cite{jimenez2021acoustic}.
It describes the porous media through five physical parameters, namely: (1) the flow resistivity $\sigma$, (2) the open porosity $\phi$, (3) the tortuosity $\alpha_{\infty}$, (4) the viscous characteristic length $\Lambda$ and (5) the thermal characteristic length $\Lambda'$.
\\
Porous materials exhibit good absorption at the medium and high frequencies, whereas their performance is somehow limited in the low-frequency regime~\cite{cao2018porous}.
To overcome this deficiency without increasing their overall thickness, specific designs of porous materials, also known as meta-porous materials or porous metamaterials, have been recently proposed~\cite{gao2022acoustic}.
The most common approaches include:
(i) introducing resonators into the porous media generating additional peaks of absorption at the desired (low) frequencies of resonance of the resonators~\cite{groby2015enhancing, lagarrigue2013absorption, zhu2019broadband};
(ii) inserting a set of rigid partitions into the porous media responsible for the formation of multiple slow waves propagating over the full layer thickness.
This approach proved to be very effective in enhancing the absorption in the low frequency regime with respect to the corresponding block of bulk porous material, though deteriorating the performances in the middle and high frequencies~\cite{yang2015metaporous, yang2016multiple};
(iii) topological optimization of the porous materials including air cavities~\cite{li2017enhanced, liu2021gradually} and micro-perforated panels~\cite{toyoda2017improved, liu2017acoustic};
(iv) multi-scale design of porous materials with slits~\cite{xin2019multiscale, attenborough2021analytical} and holes~\cite{atalla2001acoustic} allowing for sound waves with longer wavelengths to enter into the material and get dissipated;
(iv) multilayered porous media (nowadays also known as meta-porous), whose overall dynamic performance strongly depends on the geometrical arrangement of the constituent elements.
\\
Initiated in the early 90's~\cite{lauriks1992acoustic, brouard1995general, dunn1986calculation}, the meta-porous approach is again gaining increasing interest into the scientific community due to the recent developments of the acoustic metamaterials~\cite{craster2012acoustic, romero2019fundamentals} and fabrication techniques~\cite{ghaffari2015design}, which have allowed for the conception of new designs more and more optimized~\cite{zhang2020engineering, roca2021multiresonant} for an efficient and broadband sound attenuation at the low frequencies using moderate amounts of absorbing materials~\cite{gao2022acoustic}.
In this context, Jimenez et al.~\cite{jimenez2016broadband} proposed an optimized chirped multilayered porous material exhibiting enhanced low frequency absorption (compared to that of the corresponding bulk porous material of the same length).
The sound absorption and transmission of the system have been theoretically analyzed, revealing unidirectional performances, given the broken geometrical symmetry introduced by the chirped design.
Optimized layers of porous materials have also shown the possibility of reaching perfect and broadband sound absorption~\cite{jimenez2018perfect}.
In this case, the performance enhancement derived from the impedance matching of the meta-porous to the incoming wave.
Almeida et al.~\cite{almeida2023low} have investigated a multilayered porous material with slit-type perforations organized according to the Cantor set, i.e., exhibiting a fractal porosity.
A broadband and efficient absorption performance in the low-frequency regime was reported.
\\
Despite the large number of meta-porous designs proposed so far, enhancing the absorption performance at low frequencies of these materials remains an open research issue.
In this context, the concept of hierarchy, borrowed from Nature, has recently emerged as a promising source of inspiration for the engineering of metamaterials with complex structural architectures leading to advanced functional properties in several research fields, from electromagnetism to elasticity and acoustics~\cite{song2016broadband, mousanezhad2015hierarchical, miniaci2018design, dal2023bioinspired}.
Hierarchical architectures, which can be defined as recursive structural patterns repeated at different length scales, are widespread in Nature being a developmental outcome of coping with evolutionary challenges and often bring to enhanced and functional-oriented properties compared to simpler structural organizations~\cite{lakes1993materials}.
Initially mainly investigated in the quasi-static domain, hierarchical architectures have recently gained increasing interest also in dynamics and in acoustic metamaterials.
For instance, Li et al.~\cite{li2023multi} showed that adding surface porosity and unit cell heterogeneity in a multi-scale structure inspired by the cuttlefish bone allows for a broadband sound absorption and a higher deformation tolerance.
A spider-web-like organisation~\cite{ruan2022wave} or hierarchical honeycomb arrangements~\cite{sun2022topological,ma2023hierarchical} revealed to improve the bandgap properties, while acoustic metamaterials made of hierarchical membranes can reach unusual transmission loss characteristics~\cite{edwards2020transmission}.
A sandwich structure with hierarchical honeycomb interior has been shown to enhance the absorption in a much wider frequency range as compared to the regular sandwich~\cite{he2022ultralight}.
Finally, labyrinthine fractal acoustic structures have recently shown increased low-frequency sound attenuation~\cite{man2021engineering} and reflection~\cite{krushynska2017spider}.
\\
Here we introduce the concept of hierarchical meta-porous materials, i.e., multilayered stacks of porous material and air, whose geometrical organizations describe a family of structures self-replicating (with or without exact self-similarity~\cite{mazzotti2023bio}) at different length scales (see Fig~\ref{fig1}).
Through analytical calculations based on the transfer matrix method, optimization algorithms (metaheuristic Greedy Randomized Adaptive Search Procedure), and finite element simulations, we demonstrate the enhancement of the absorption coefficient as additional hierarchical levels are added into the porous layers(s).
Two types of organization are investigated, and their performance compared to that of the corresponding block of bulk material: a hierarchical periodic (HP) and a hierarchical gradient (HG).
In both cases it is found that the introduction of multiple levels of hierarchy simultaneously allows to (i) increase the level of absorption, and (ii) reduce the quantity of porous material required to achieve better performance.
Both the cases of normal and oblique incidences are examined.
The optimization procedure has been applied to maximize the absorption of the highest hierarchical level in the desired frequency range ([20, 2000] Hz), since for each proposed structure adding further hierarchical levels is always possible.
\\
The paper is organized as follows: Section~\ref{Mam} is devoted to the description of the models and methods.
First, the hierarchical periodic and gradient periodic designs are introduced.
Then, the type of model adopted to describe the porous layers and the formalism used to calculate their absorption properties are provided.
Finally, the details of the optimization algorithm exploited to define the designs are given.
Section~\ref{Res} reports the results for the cases of normal and oblique plane wave incidence, as well as the impact of the flow resistivity on the absorption coefficients.
Conclusions and further perspectives are given in Section~\ref{Conclusions}.

\section{Models and methods}
\label{Mam}
\subsection{Hierarchical periodic and hierarchical gradient design}
\label{Hpahgd}
A great variety of biological systems exhibit a hierarchical organization, i.e., recursive geometrical patterns repeating at the micro- and/or macro-scale~\cite{fratzl2007nature} scales.
Hierarchy often derives from heterogeneity introduced in the form of (i) reinforcing elements (fibers, platelets, or crystals embedded in a softer matrix), (ii) voids, cavities, or canals into a matrix, and/or (iii) alternating layers of stiffer and softer materials.
These diverse organizations can contribute to increased energy dissipation and crack deflection capabilities, toughness, resilience, but also to potentially affect the propagation of elastic waves or damping along with an overall mass density reduction~\cite{miniaci2018design}.
Indeed, the evidence for a strong correlation between the propagation of waves and the micro-structure of the bio-composites hosting the propagation has been shown, connecting it to the hierarchical organization, often consisting of hard material building blocks embedded in a soft organic matrix, assembled in a hierarchical manner across multiple length scales~\cite{zhang2013broadband}.
Hierarchy can occur at different length scales, as in bones, nacre, or, at similar length scales, as in leaves, wood, corals, or sponges (porous materials)~\cite{petkovich2012hierarchically}.
Electron microscope images of such hierarchical structures have often shown that their micro-structure appears not only to be self-similar but also periodic at each level of their structural hierarchy, such as in the case of enamel and dentine, protecting our teeth from failure after millions of times of mastication, or lobster cuticles, shrimp clubs, and crab claws exhibiting exceptional resistance to repeated dynamic attack either for preying or shielding purposes~\cite{weaver2012stomatopod}.
\begin{figure*}[ht!]
\centerline{\includegraphics[width=5in]{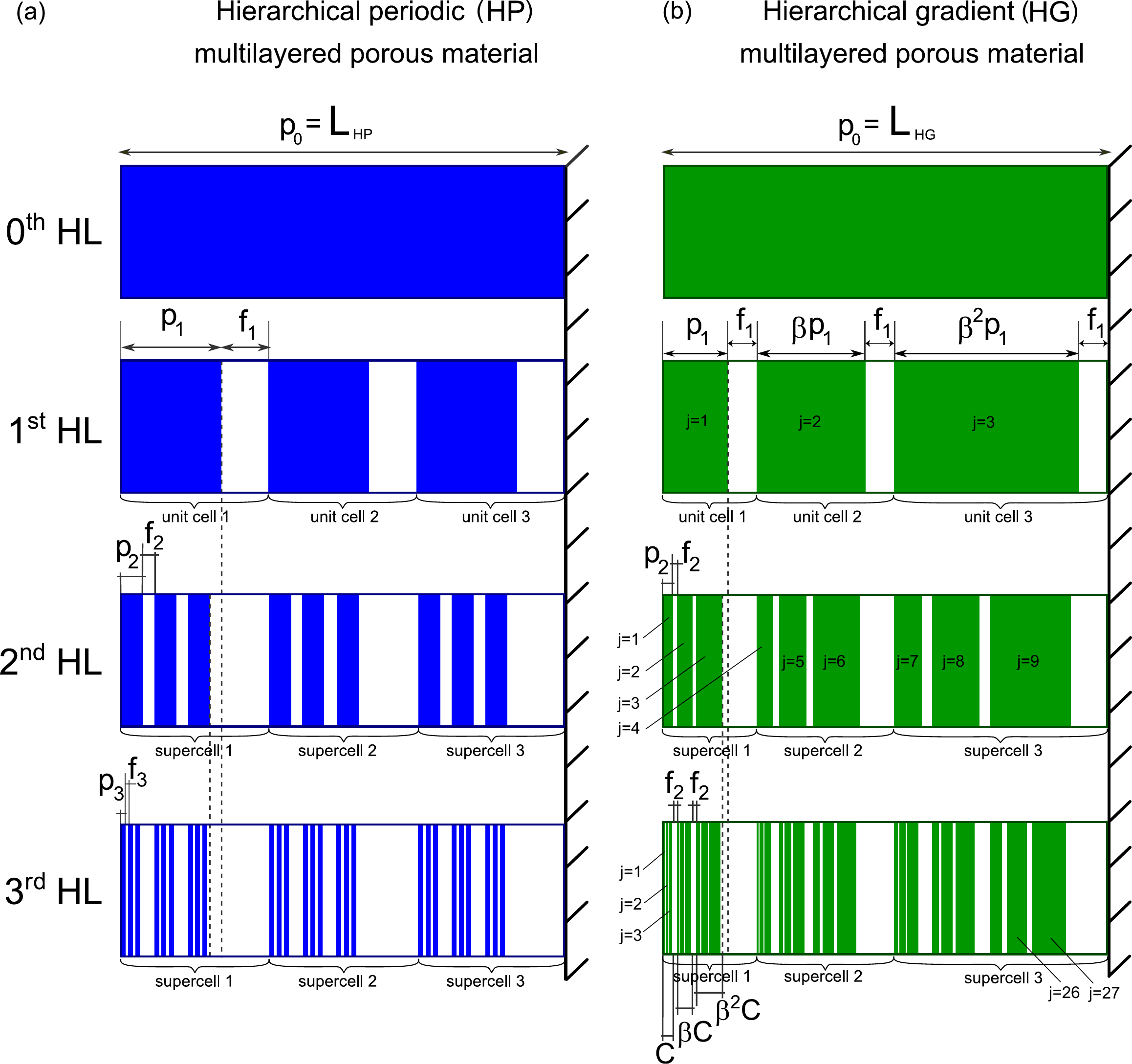}}
\caption{\textbf{Hierarchical periodic and hierarchical gradient multilayered porous structures}.
(a) A hierarchical periodic (HP) multilayered porous structure.
The porous layers are colored in blue, and the air gaps in white.
(b) A hierarchical gradient (HG) multilayered porous structure.
The porous layers are colored in green, and the air gaps in white.
In both cases, $m=3$ hierarchical levels are investigated, and the simplest organization, the $0^{\rm{th}}$ hierarchical level, reduces to a bulk block of porous material.
The recursive rule to build each next hierarchical levels $i^{\rm{th}}$, with $i \in [1,m]$, is to divide the porous domain(s) of the previous level $(i-1)^{\rm{th}}$ of length $p_{i-1}$ in $n=3$ porous and air layers.
In (a) the resulting layers have constant lengths $p_{i}$ and $f_{i}$ throughout the sample, while in (b) the structure self-replicates following the scaling rule $p_i^{(j)} = p_i\beta^{j-1-(n-1)\sum\limits_{q = 1}^{i-1}\left[\frac{j-1}{n^q}\right]}$, being $\beta$ a constant parameter, $i \in [0, m]$ the considered HL, $j \in [1, n \times m]$ the porous layer sequence number, $p_i$ the length of the first porous layer of the $i^{\rm{th}}$ HL and $[ \cdot ]$ indicates the integer part of the number included within the square brackets.}
\label{fig1}
\raggedright
\end{figure*}
\\
Inspired by these patterns, here we propose and investigate a multilayered porous media with a (i) hierarchical periodic (HP) and (ii) hierarchical gradient (HG) organization, as reported in Fig.~\ref{fig1}(a) and Fig.~\ref{fig1}(b), respectively.
In our case, the stiffer layers (colored in blue for the HP and in green for the HG) are made of porous material (melamine), whereas the softer ones (in white) of air.
In both cases, $m=3$ hierarchical levels (HL) are investigated, and the simplest organization, the $0^{\rm{th}}$ hierarchical level, reduces to a bulk block of porous material of length $p_0$ (top panels of Fig.~\ref{fig1}), where $p_0 = L_{\rm{HP}} = 26$ cm in the case of HP, and $ p_0 = L_{\rm{HG}} = 41.3$ cm in the case of HG.
\\
In the HP study case, the recursive rule adopted to build each next hierarchical level $i^{\rm{th}}$, with $i \in [1,m]$, is to divide the porous domain(s) of the previous level $(i-1)^{\rm{th}}$ of length $p_{i-1}$ in $n=3$ unit cells (supercells) made of porous and air layers of length $p_{i}$ and $f_{i}$, respectively.
For instance, the 1$^{\rm{st}}$ hierarchical level is made of three unit cells of alternating porous and air layers of length $p_1$ and $f_1$, such that $ n \cdot ( p_1 + f_1 ) = p_0$.
In the same manner, the 2$^{\rm{nd}}$ hierarchical level divides each porous domain of length $p_1$ of the 1$^{\rm{st}}$ hierarchical level into three unit cells of alternating porous and air layers of length $p_2$ and $f_2$, respectively, so that they form three supercells of total length $n\cdot(p_2+f_2)+f_1 = p_1+f_1$.
Finally, the supercell of the last hierarchical level (3$^{\rm{rd}}$ HL) consists of the following combination of alternating porous and air layers
\textcolor{green}{$p_3f_3p_3f_3p_3f_3$}
\textcolor{blue}{$f_2$}
\textcolor{green}{$p_3f_3p_3f_3p_3f_3$}
\textcolor{blue}{$f_2$}
\textcolor{green}{$p_3f_3p_3f_3p_3f_3$}
\textcolor{blue}{$f_2$}
\textcolor{red}{$f_1$}.
The parameters $m$ and $n$ have been here arbitrarily chosen to be equal to 3 to maintain a good compromise between a sufficient high number of hierarchical levels explored and a limited total length of the structure.
Nevertheless, similar reasoning can be applied for higher or lower values of $n$.
The thinnest porous layer thickness has been limited to at least 4 mm so that the JCA model properly describe its behaviour -- see subsection~\ref{Motpm}.
The lengths of the porous and fluid layers are determined as a result of an optimization procedure applied to the last hierarchical level -- see Section~\ref{Ootg}.
\\
In the HG study case, the finite structures of each HL (other than the $0^{\rm{th}}$ HL, which is a bulk block of porous material) are made of $n = 3$ unit cells (supercells) constituted of porous and air layers, as in the HP case.
The difference, here, is that the three unit cells (supercells) belonging to the same HL exhibit different lengths according to the rule
\begin{equation}
p_i^{(j)} = p_i\beta^{j-1-(n-1)\sum\limits_{q = 1}^{i-1}\left[\frac{j-1}{n^q}\right]},
\end{equation}
where $\beta$ is a constant parameter (determined to be equal to 1.1085 by the optimization algorithm -- see Section~\ref{Ootg}), $i \in [0, m]$ indicates the HL, $j \in [1, n \times m]$ the porous layer sequence number, $p_i$ the length of the first porous layer of the $i^{\rm{th}}$ HL and $[ \cdot ]$ indicates the integer part of the number included within the square brackets.
The recursive rule to build each next hierarchical level $i^{\rm{th}}$, with $i \in [1,m]$, is to divide the porous domain(s) of the previous $(i-1)^{\rm{th}}$ HL into $n$ porous and air layers keeping the proportions of the previous HL constants, i.e., maintaining the ratio $\frac{p_i}{f_i}$ constant.
More qualitatively, the process can be described as taking all the porous and fluid layers of the whole structure of the $(i-1)^{\rm{th}}$ HL and squeeze them successively into the lengths corresponding to those of the $j^{\rm{th}}$ porous layer(s) of the 1$^{\rm{st}}$ HL (minus the length(s) of the last fluid layer(s) of the 2$^{\rm{nd}}$ to the $(i-1)^{\rm{th}}$ HL(s)) to form the $(j-1)n^{i-1}+1$ up to $jn^{i-1}$ porous layers of the successive $i^{\rm{th}}$ HL (in this recursive formula the indices $j$ always refer to the $1^{\rm{st}}$ HL).
For instance, the 1$^{\rm{st}}$ hierarchical level is made of $n = 3$ unit cells of alternating porous and air layers of lengths $p_1,f_1,\beta p_1, f_1, \beta^2p_1, f_1$, such that $p_1 \cdot \sum\limits_{i = 1}^n \beta^{i-1}+ n \cdot f_1  = p_0$.
In the same manner, the 2$^{\rm{nd}}$ hierarchical level divides each porous domain $j^{\rm{th}}$ of the $1^{\rm{st}}$ hierarchical level into $n = 3$ alternating porous and air layers of lengths $p_2^{(t)}$ and $f_2^{(t)}$, respectively, so that $\sum\limits_{t = 1}^n (p_2^{(t)}+f_2^{(t)})= p_1^{(j)}$.
Finally, the last HL (the 3$^{\rm{rd}}$ one) appears to be constructed from the supercells $\textcolor{purple}{C} = \textcolor{green}{p_3},\textcolor{green}{f_3},\beta \textcolor{green}{p_3},\textcolor{green}{f_3},\beta^2 \textcolor{green}{p_3}, \textcolor{green}{f_3}$ in the following sequence
\begin{eqnarray}
\textcolor{green}{C},\textcolor{blue}{f_2},\beta\textcolor{green}{C,}\textcolor{blue}{f_2},\beta^2\textcolor{green}{C},\textcolor{blue}{f_2},\textcolor{red}{f_1},\\  \nonumber
\beta\textcolor{green}{C},\beta\textcolor{blue}{f_2,}\beta^2\textcolor{green}{C},\beta\textcolor{blue}{f_2},\beta^3\textcolor{green}{C},\beta\textcolor{blue}{f_2},\textcolor{red}{f_1},\\ \nonumber
\beta^2\textcolor{green}{C},\beta^2\textcolor{blue}{f_2},\beta^3\textcolor{green}{C},\beta^2\textcolor{blue}{f_2},\beta^4\textcolor{green}{C},\beta^2\textcolor{blue}{f_2},\textcolor{red}{f_1}, \nonumber
\end{eqnarray}
and how is illustrated in Fig.~\ref{fig1}(b).

\subsection{Modeling of the porous material}
\label{Motpm}
The porous layers are made of melamine, whose propagation of sound is described through the Johnson-Champoux-Allard (JCA) model~\cite{allard2009propagation}.
The following parameters are adopted:
porosity $\phi = 0.98$,
tortuosity $\alpha_{\infty} = 1.34$,
viscous characteristic length $\Lambda = 150$ $\mu$m,
thermal characteristic length $\Lambda^{\prime} = 560$ $\mu$m, and
static air-flow resistivity $\sigma = 49000$ Pa.s/m$^2$ and $\sigma = 10000$ Pa.s/m$^2$ (chosen because of being representative of rather common porous media).
The effective density and bulk modulus of the porous material are expressed as~\cite{allard2009propagation, de2007acoustic}
\begin{align}
\begin{split}
\rho_p = \frac{\rho_0 \alpha_{\infty}}{\phi}\left(1-\frac{i\omega_c}{\omega}F(\omega) \right),\\
K_p = \frac{\gamma P_0}{\phi}\left(\gamma-(\gamma-1)\left(1-\frac{i \omega_c^{\prime}}{\rm{Pr} \omega} G(\rm{Pr} \omega)  \right)^{-1} \right)^{-1},
\end{split}
\end{align}
where $\omega_c = \frac{\sigma\phi}{\rho_0 \alpha_{\infty}}$ is the Biot frequency, $\omega_c^{\prime} = \frac{\sigma^{\prime}\phi}{\rho_0 \alpha_{\infty}}$, $\sigma^{\prime} = \frac{8\alpha_{\infty}\eta}{\phi \Lambda^{\prime 2}}$~\cite{champoux1991dynamic} is the thermal resistivity~\cite{groby2011enhancing} and the correction functions are given by~\cite{allard1992new,johnson1987theory}
\begin{align}
\begin{split}
F(\omega) = \sqrt{1+i\eta \rho_0 \omega\left(\frac{2\alpha_{\infty}} {\sigma \Lambda\phi}\right)^2},\\
G(Pr \omega) = \sqrt{1+i\eta\rho_0\rm{Pr}\omega \left(\frac{2\alpha_{\infty}} {\sigma^{\prime} \Lambda^{\prime}\phi}\right)^2},
\end{split}
\end{align}
where $P_0 = 101.325$ kPa is the atmospheric pressure, and the parameters of air (which is assumed to be the fluid present in the pores) are given by: $\rho_0 = 1.204$ kg/m$^3$ (density of the air), Pr $= 0.71$ (Prandtl number), $\gamma = 1.4$ (heat capacity ratio) and $\eta = 1.839 \cdot 10^{-5}$ kg/(ms) (air viscosity).
Once these parameters determined, the wavenumber $k_p = \omega \sqrt{\frac{\rho_p}{K_p}}$ and the effective acoustic impedance $Z_p = \sqrt{\rho_p K_p}$ can be obtained.

\subsection{Calculation of the absorption coefficient}
\label{Cotac}
A harmonic plane wave with time dependence $e^{-i\omega t}$ incident on the rigidly backed multilayered hierarchical structures is considered.
The transfer matrix formalism is adopted.
The transfer matrix of each layer can be defined as
\begin{equation}
\label{EqRef}
T_i = 
\begin{pmatrix} 
\cos{k_{ix} l_i} & \frac{iZ_i}{\cos{\theta}}\sin{k_{ix} l_i}\\
\frac{i \cos{\theta}}{Z_i}\sin{k_{ix} l_i} & \cos{k_{ix} l_i}
\end{pmatrix},
\end{equation}
where $l_i$ is the length of the corresponding layer, $k_{ix} = k_{px} (k_{0x})$ is the projection of the wavenumber perpendicular to the layers, $Z_i = Z_p(Z_0)$ is the impedance, $\theta$ is the angle of the incident plane wave considered.
The subscripts $_p$ and $_0$ indicate if the property refers to the porous or fluid (air) layer, respectively.
The total transfer matrix $T^{(n)}$ of the $n^{\rm{th}}$ hierarchical level is the consecutive product of the transfer matrices of each layer 
\begin{equation}
T^{(n)} = \prod_{i} T_i.
\end{equation}
The reflection coefficient is obtained as follows
\begin{equation}
R_n = \frac{T^{(n)}_{11}\cos{\theta}-Z_0 T^{(n)}_{21}}{T^{(n)}_{11}\cos{\theta}+Z_0 T^{(n)}_{21}},
\end{equation}
and the absorption determined as $\alpha_n = 1-|R_n|^2$.

\subsection{Optimization of the geometry}
\label{Ootg}
For the sake of comparison of the absorption performances of the multilayered structures as additional hierarchical levels are introduced, the absorption coefficient of the highest hierarchical level (3$^{\rm{rd}}$ one in our case) is optimized by minimizing the area above the absorption curve $\alpha_3(f)$ in the desired frequency range [20, 2000] Hz.
This strategy has been chosen because for a given multilayered structure additional hierarchical levels can always be added.
Once the geometrical parameters of the highest hierarchical level are determined (lengths $p_i$ and $f_i$ of the porous and air layers), those of the previous levels can be recursively derived -- see Section~\ref{Hpahgd}.
\\
The optimization algorithm adopted is based on the resolution scheme of the metaheuristic Greedy Randomized Adaptive Search Procedure (GRASP~\cite{feo1995greedy}).
A solution is constructed, without backtracking, in a very short time (between 1.8 and 2.0 s).
According to the type of structure to optimize (HP or HG), a large variety of solutions is obtained through a randomized exploration of the greedy algorithm which assigns values to the variables describing the lengths of the porous ($p_3$ in the HP and $p_3^j$ in the HG) and fluid ($f_i$) layers, the number of hierarchical levels $m$, the number of unit cells (supercells) $n$, and/or the evolution coefficient $\beta$ (this last one concerning the HG study case, only).
Restricted by practical requirements, the aforementioned parameters were bounded to $p_i \ge 4$ mm, $f_i \ge 1.8$ mm, $m \le 3$, $n \le 3$ in the optimization procedure.
The obtained solutions are then sorted based on a $\ll$best absorption$\gg$ criterion, which forms the objective function.
\\
The final parameters for the HP organization are: $L_{HP} = 26$ cm, $p_3 = 4$ mm, $f_3 = 2$ mm, $f_2 = 6$ mm, $f_1 = 14.7$ mm.
Those for the HG multilayered structure are: $L_{HG} = 41.3$ cm $p_3 = 4$ mm, $f_3 = 1.8$ mm, $f_2 = 8.4$ mm, $f_1 = 39.4$ mm and $\beta = 1.1085$.
The sizes are rounded up to $0.1$ mm.

\section{Results}
\label{Res}
\subsection{Periodic hierarchical material}
\subsubsection*{Normal incidence}
\begin{figure*}[ht!]
\centerline{\includegraphics[width=5 in]{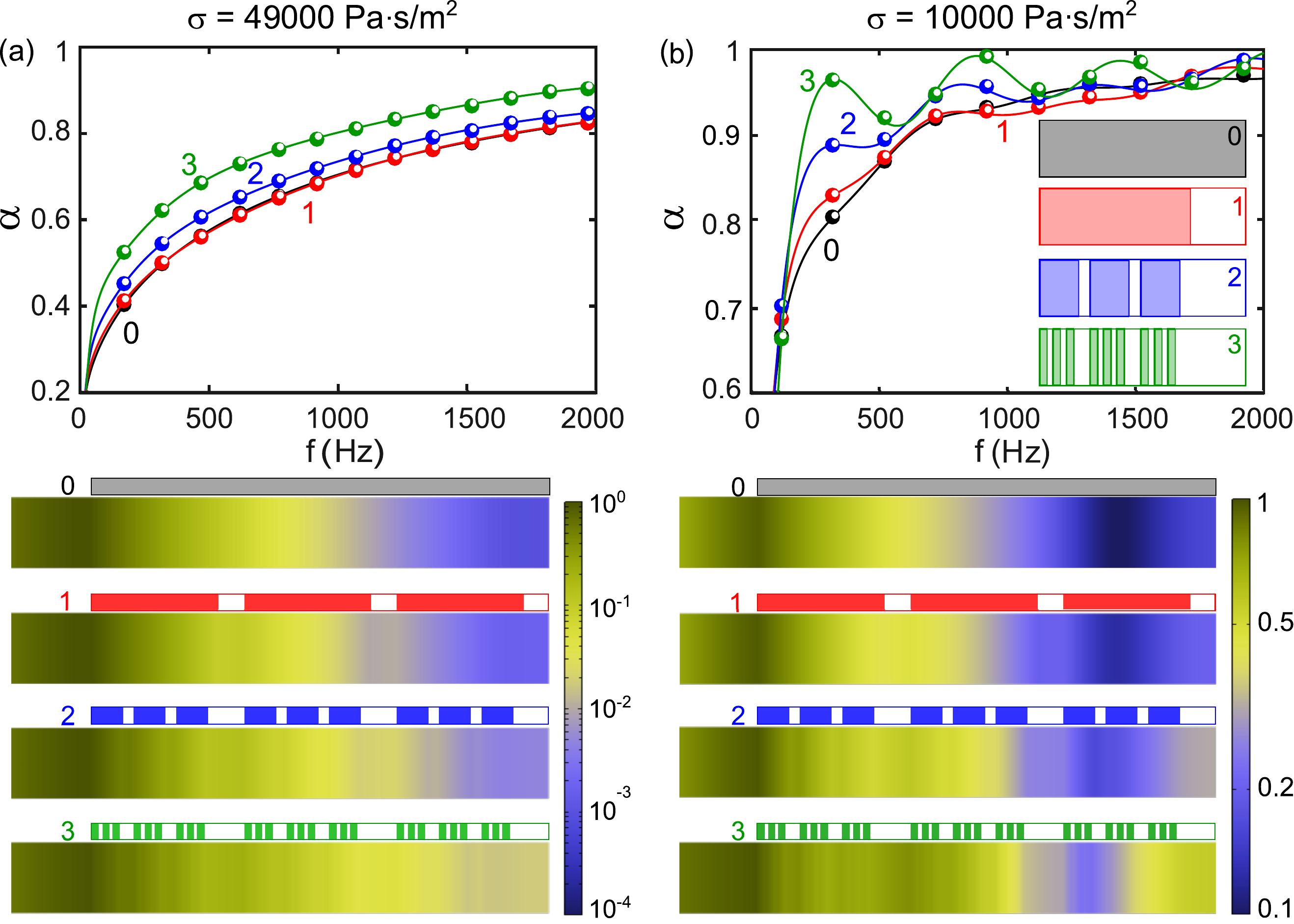}}
\caption{\textbf{Absorption coefficient and normalized total acoustic pressure fields at normal incidence}.
Top panels: absorption coefficient $\alpha$ for different hierarchical levels when (a) $\sigma$ = 49000 Pa.s/m$^2$ and (b) $\sigma$ = 10000 Pa.s/m$^2$.
Lower panels: normalized total pressure fields $\frac{|p|}{\max{|p|}}$ reconstructed at $f$ = 900 Hz.
The disposal of the porous layers with respect to the whole multilayered structures is plotted above each pressure distribution.}
\raggedright
\label{fig2}
\end{figure*}

Figure~\ref{fig2} reports the absorption coefficient $\alpha$ as a function of the frequency $f$ (top panels) and the normalized total pressure fields at 900 Hz (bottom panels) for the HP study case.
The comparison includes three hierarchical levels besides the case of bulk block of porous material (0$^{\rm{th}}$ HL) for different values of $\sigma$ (49000 Pa.s/m$^2$ in Fig.~\ref{fig2}(a) and 10000 Pa.s/m$^2$ in Fig.~\ref{fig2}(b), respectively).
The sketch of a unit cell / supercell of each hierarchical level is shown in the inset of Fig.~\ref{fig2}(b).
A plane wave normally incident from the left side of the structures is considered.
The results, issued from the transfer matrix method described in the previous Sections~\ref{Motpm} and~\ref{Cotac}, are reported as continuous colored lines (black for the 0$^{\rm{th}}$ HL, red for the 1$^{\rm{st}}$ HL, blue for the 2$^{\rm{nd}}$ HL, and green for the 3$^{\rm{rd}}$ HL).
The absorption coefficients are calculated also numerically through finite element methods using the $\ll$Acoustics$\gg$ module of Comsol Multi-physics, and reported as colored circular markers (also in this case in black for the 0$^{\rm{th}}$ HL, red for the 1$^{\rm{st}}$ HL, blue for the 2$^{\rm{nd}}$ HL, and green for the 3$^{\rm{rd}}$ HL).
A perfect agreement is found over the whole frequency range [20 2000] Hz.
\\
Top panels of Fig.~\ref{fig2}(a) and Fig.~~\ref{fig2}(b) clearly show that the introduction of hierarchy brings to a considerable enhancement of the absorption properties of the structure, regardless the chosen value of $\sigma$ (49000  Pa.s/m$^2$ or 10000  Pa.s/m$^2$), but for the first iteration (0$^{\rm{th}}$ HL $\Longrightarrow$ 1$^{\rm{st}}$ HL), where the variation of $\alpha$ is minimal.
The introduction of additional HLs in porous media with large values of $\sigma$, top panel of Fig.~\ref{fig2}(a), brings to a progressive increase of the absorption coefficient (going up to a maximum enhancement of 39\% at $f$ = 75 Hz and between 20\% and 25\% when $f \ge$ 200 Hz when the 3$^{\rm{rd}}$ HL is compared to the bulk block of porous material -- 0$^{\rm{th}}$ HL) over the whole considered frequency range.
Similarly, when smaller values of $\sigma$ are considered, top panel of Fig.~\ref{fig2}(b), the introduction of hierarchy still remains advantageous, especially at the low frequencies, although the enhancement of the absorption coefficient has now an oscillating behaviour reaching a 20\% enhancement at $f$ = 330 Hz (where an absorption peak deriving from the $\lambda/4$ resonance is expected).
The \% of absorption enhancement reduces going towards higher frequencies.
\\
Finally, to get further insight on the possible reason producing the observed increase of the absorption coefficient as additional hierarchical levels are added, the normalized total acoustic pressure fields $\frac{|p|}{\max{|p|}}$ for the 3$^{\rm{rd}}$ HL at $f$ = 900 Hz (where the most pronounced peak of the green curve in Fig.~\ref{fig2}(b) occurs) are reported in the lower panels of Fig.~\ref{fig2}.
The disposal of the porous layers with respect to the whole multilayered structures is plotted above each pressure distribution.
From these plots it clearly emerges that introducing additional hierarchical levels allows the pressure field to propagate further inside the hierarchical meta-porous, which in turn can explain the higher values of $\alpha$.
The main physical mechanism taking place seems to be related to the fact that hierarchy, alternating porous and air layers at different length scales, increases its absorption efficiency by $\ll$trapping$\gg$ the waves inside the air gaps.
Specifically, in the case of smaller value of $\sigma$, the introduction of additional hierarchical levels allows the incident field to propagate further within the sample and to increase the absorption values related to the first quarter-wavelength resonance (around 330 Hz) and to the higher order resonances -- Fig.~\ref{fig2}(b).
On the contrary, when larger values of $\sigma$ are considered, the field decreases rapidly inside the structure becoming very weak at the rigid end and not allowing for any noticeable geometrical resonances -- Fig.~\ref{fig2}(a).
\subsubsection*{Oblique incidence}
\begin{figure*}[ht!]
\centerline{\includegraphics[width=5 in]{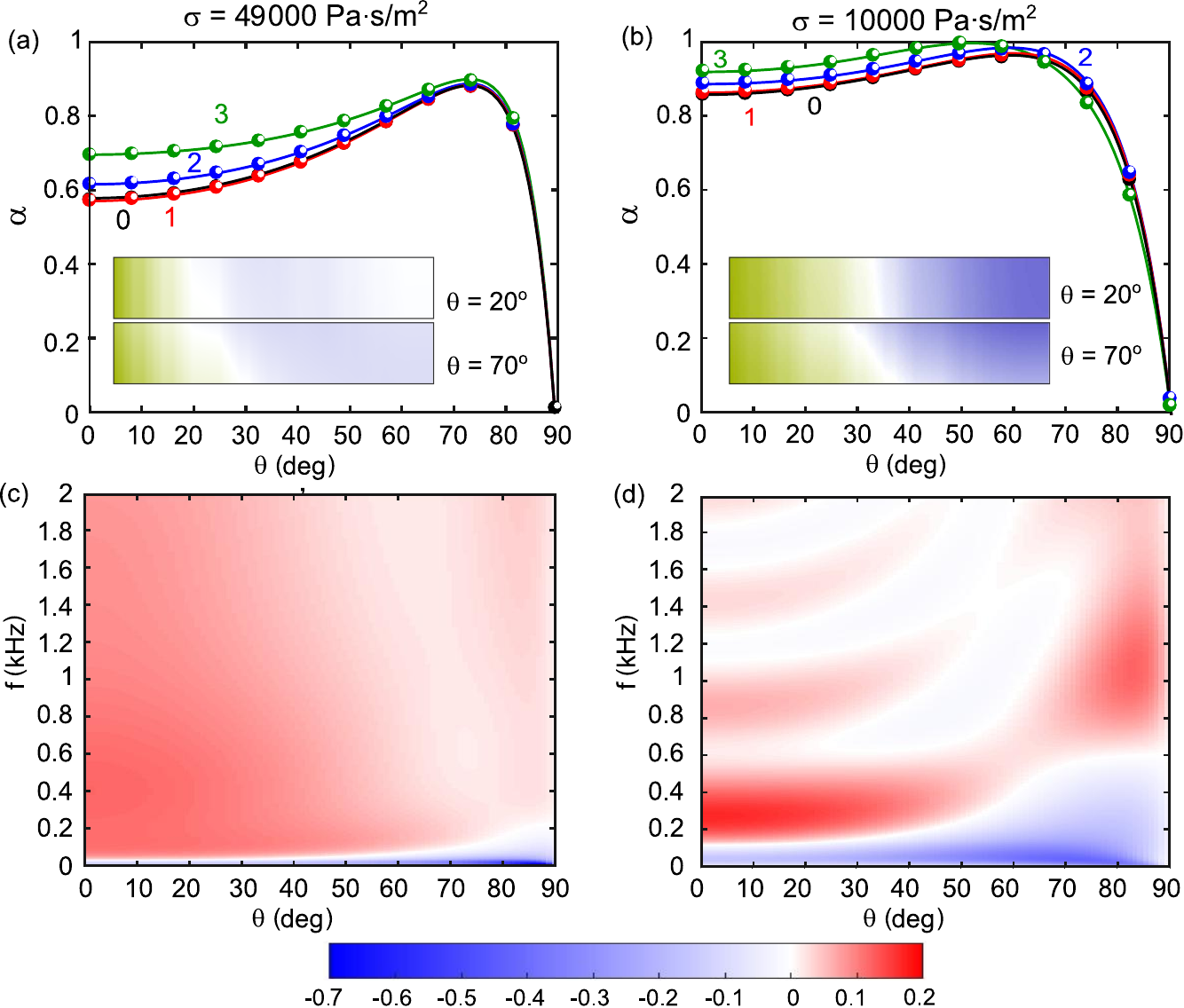}}
\caption{\textbf{Influence of the angle on the absorption coefficient}.
(a, b) The absorption coefficient $\alpha$ at 500 Hz as a function of $\theta \in$ [0, 90]$^\circ$ for the three hierarchical levels and the bulk block of porous material (0$^{\rm{th}}$ HL).
Different values of $\sigma$ (49000 Pa.s/m$^2$ and 10000 Pa.s/m$^2$) are considered.
The normalized total acoustic pressure fields $\frac{\Re{(p)}}{\max{\Re{(p)}}}$ at different angles of incidence ($\theta = 20^\circ$ and $\theta = 70^\circ$) are reported as insets.
(c, d) The maps of the difference between the absorption coefficient $\alpha_3$ of the 3$^{\rm{rd}}$ HL and $\alpha_0$ of the 0$^{\rm{th}}$ HL in the $(f,\theta)$ space for $\sigma$ = 49000 Pa.s/m$^2$ and 10000 Pa.s/m$^2$, respectively.}
\raggedright
\label{fig3}
\end{figure*}

The absorption performance depends on the angle of incidence $\theta$ of the plane wave impinging the hierarchical meta-porous under consideration, as clearly evident from Eq.~\ref{EqRef}.
Figure~\ref{fig3}(a) and Fig.~\ref{fig3}(b) report the absorption coefficient $\alpha$ at 500 Hz as a function of $\theta \in$ [0, 90]$^\circ$ for the three hierarchical levels and the bulk block of porous material.
Also in this case, two values of $\sigma$ are considered (49000 Pa.s/m$^2$ and 10000 Pa.s/m$^2$, respectively).
\\
The results, issued from the transfer matrix method (continuous colored lines) and finite element-based numerical simulations (circular colored markers), clearly show that the hierarchical design is advantageous, in terms of absorption, for most of the angles of incidence, namely $\theta \in [0^\circ, 75^\circ]$ when $\sigma$ = 49000 Pa.s/m$^2$ and $\theta \in [0^\circ, 65^\circ]$ when $\sigma$ = 10000 Pa.s/m$^2$.
For larger angles of incidence, the absorption curves almost merge and go to 0 when $\theta = 90^\circ$.
The largest enhancement between the bulk block of porous material (0$^{\rm{th}}$ HL) and the highest HL considered (3$^{\rm{rd}}$ HL) is observed at normal incidence ($\theta = 0$), for both the values of $\sigma$, although it reaches 20\% and 10\% for the larger and smaller values of $\sigma$, respectively.
It is worth reminding here that the geometry was only optimized for normal incidence, i.e., $\theta = 0^\circ$, which, in our opinion, makes the potential of the hierarchical design remarkable since it remains advantageous also for most of the remaining angles.
The normalized total acoustic pressure fields $\frac{\Re{(p)}}{\max{\Re{(p)}}}$ for the 3$^{\rm{rd}}$ HL at different angles of incidence ($\theta = 20^\circ$ and $\theta = 70^\circ$) are also reported as insets in Fig.~\ref{fig3}(a) and Fig.~\ref{fig3}(b).
The pressure levels reach much higher values inside the structures composed of porous with lower values of $\sigma$, allowing for the more efficient absorption.
\\
What happens in the remaining frequencies can be deduced from Fig.~\ref{fig3}(c) and Fig.~\ref{fig3}(d), which report the maps of the difference between the absorption coefficient $\alpha_3$ of the 3$^{\rm{rd}}$ HL and $\alpha_0$ of the 0$^{\rm{th}}$ HL in the $(f,\theta)$ space for $\sigma$ = 49000 Pa.s/m$^2$ and 10000 Pa.s/m$^2$, respectively.
The color map has been chosen so that the regions where no enhancement of the absorption was observed (when the hierarchical design is introduced), appear in white, whereas the regions where the hierarchy outperforms (under-performs) the bulk block porous layer are in dark red (blue).
In this sense larger values of $\sigma$ seem to be more beneficial in the overall range of parameters, though for certain regions the smaller value of $\sigma$ provides larger values of $\alpha_3-\alpha_0$.

\subsubsection*{Influence of the flow resistivity}
\begin{figure*}[ht!]
\centerline{\includegraphics[width= 5 in]{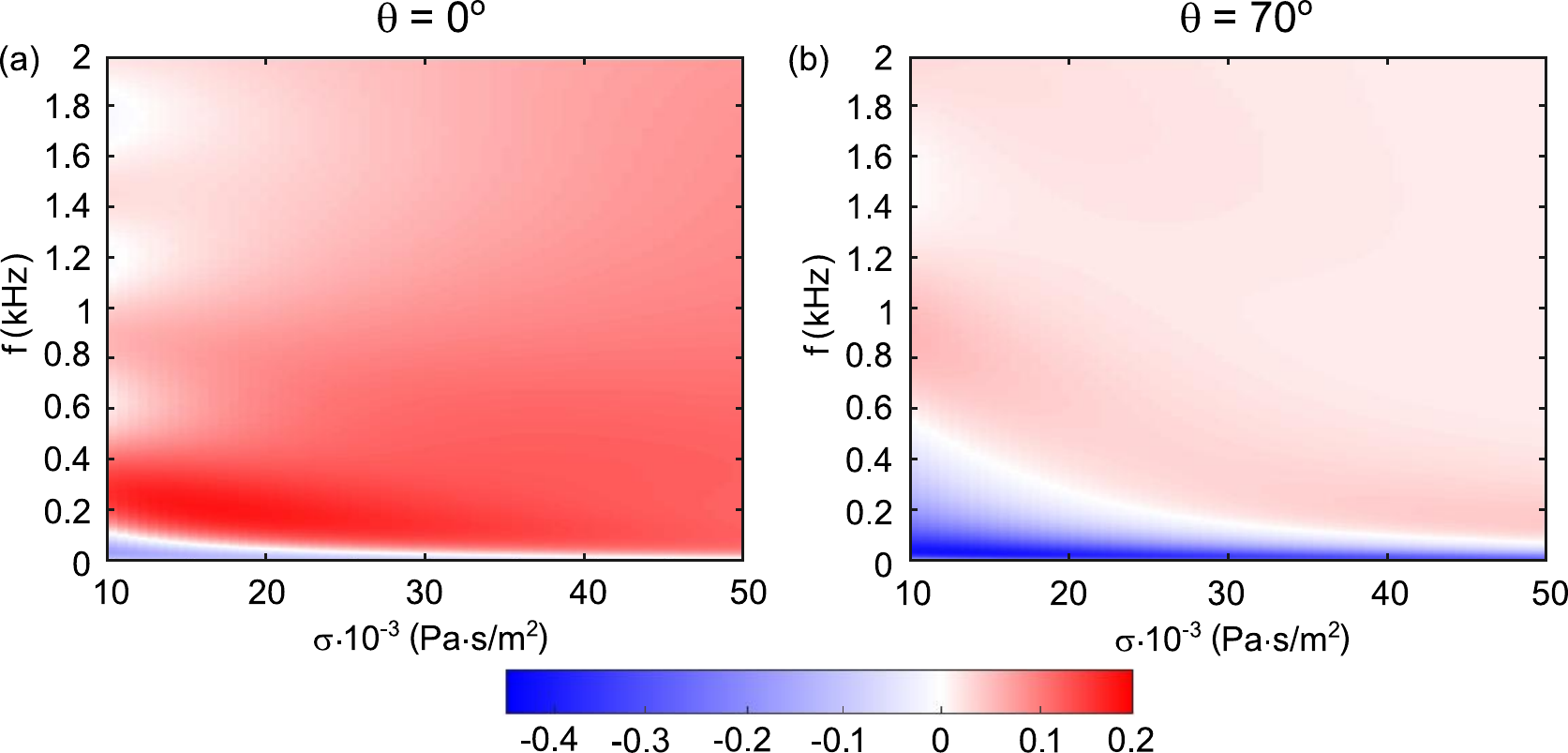}}
\caption{\textbf{Influence of the flow resistivity on the absorption coefficient}.
Difference $\alpha_3-\alpha_0$ as a function of $f$ and $\sigma$ for (a) $\theta = 0^{\circ}$ and (b) $\theta = 70^{\circ}$.}
\raggedright
\label{fig4}
\end{figure*}

Finally, the behaviour of $\alpha_3-\alpha_0$ is reported in Fig.~\ref{fig4}(a) and Fig.~\ref{fig4}(b) as a function of the frequency $f$ and of the flow resistivity $\sigma$, when $\theta = 0^{\circ}$ and $\theta = 70^{\circ}$ (the range 10000 - 50000 Pa.s/m$^2$ for $\sigma$ has been considered since it corresponds to common values of porous media used in practical applications).
The same color map described above applies.
Therefore, we can infer that, when normal incidence is considered, Fig.~\ref{fig4}(a), the enhancement of the absorption coefficient introduced by the hierarchical design (3$^{\rm{rd}}$ HL compared to the bulk block of porous material) increases as the value of $\sigma$ increases (except a narrow low frequency region).
When $\theta$ = 70$^\circ$ is considered, Fig.~\ref{fig4}(b), the frequency region within which the hierarchical design under-performs the bulk block of porous material slightly increases, especially at the lower frequencies.

\subsection{Gradient hierarchical material}
\subsubsection*{Normal incidence}
\begin{figure*}[ht!]
\centerline{\includegraphics[width=5 in]{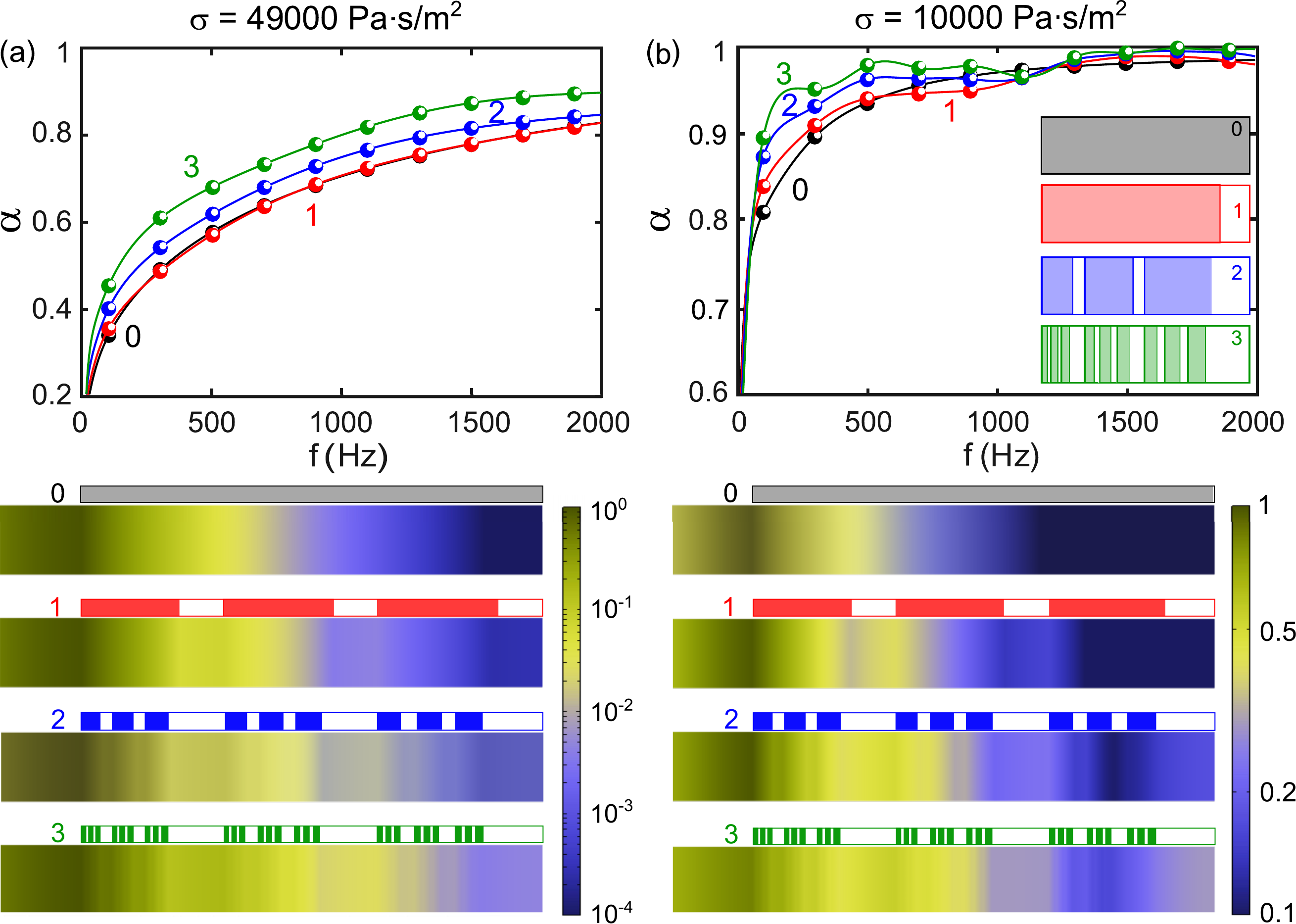}}
\caption{\textbf{Absorption coefficient and normalized total acoustic pressure fields at normal incidence}.
Top panels: absorption coefficient $\alpha$ for different hierarchical levels when (a) $\sigma$ = 49000 Pa.s/m$^2$ and (b) $\sigma$ = 10000 Pa.s/m$^2$.
Lower panels: normalized total pressure fields $\frac{|p|}{\max{|p|}}$ reconstructed at $f$ = 600 Hz.
The disposal of the porous layers with respect to the whole multilayered structures is plotted above each pressure distribution.}
\raggedright
\label{fig5}
\end{figure*}

Following the same approach described in the previous Section, meta-porous with a hierarchical gradient (HG) organization, as the ones reported in Fig.~\ref{fig1}(b), are considered.
First, the absorption coefficient $\alpha$ as a function of the frequency $f$ and the normalized total pressure fields at 600 Hz are reported in the top and bottom panels of Fig.~\ref{fig5}, respectively.
As in the HP case, three HLs and the bulk block of porous material are compared when $\sigma$ = 49000 Pa.s/m$^2$, 
Fig.~\ref{fig5}(a), and when $\sigma$ = 10000 Pa.s/m$^2$, Fig.~\ref{fig5}(b).
The results are qualitatively and quantitatively very similar to those of the HP organization (compare Fig.~\ref{fig2} and Fig.~\ref{fig5}).
In both plots the absorption coefficient benefits from the introduction of hierarchy, though its oscillating behaviour when $\sigma$ = 10000 Pa.s/m$^2$, Fig.~\ref{fig5}(b), is smoother.
The total pressure field maps $\frac{|p|}{\max{|p|}}$ at $f = 600$ Hz, where we observe a peak of the green curve in Fig.~\ref{fig5}, are reported in the lower panel.
Increasing the number of HLs, the wave field propagates over a longer distance inside the material, and, thus, the wave interacts with multiple porous layers to get highly absorbed.
When $\sigma$ = 10000 Pa.s/m$^2$, the quarter-wavelength (around 200 Hz) and higher order resonances take place.
Also in this case, the hierarchy seems to get these peaks more pronounced.

\subsubsection*{Oblique incidence}
\begin{figure*}[ht!]
\centerline{\includegraphics[width=5 in]{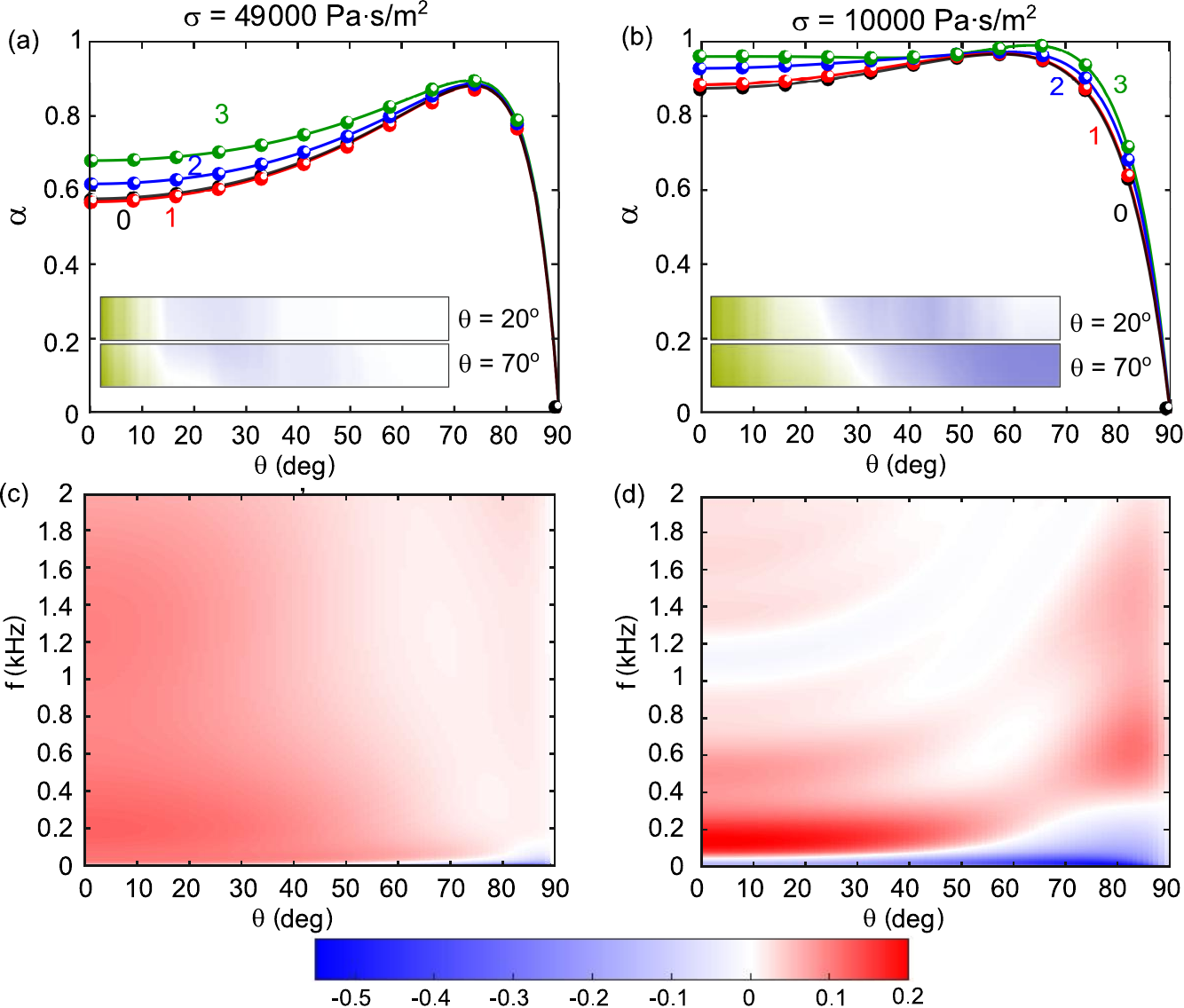}}
\caption{\textbf{Influence of the angle on the absorption coefficient}.
(a, b) The absorption coefficient $\alpha$ at 500 Hz as a function of $\theta \in$ [0, 90]$^\circ$ for the three hierarchical levels and the bulk block of porous material (0$^{\rm{th}}$ HL).
Different values of $\sigma$ (49000 Pa.s/m$^2$ and 10000 Pa.s/m$^2$) are considered.
The normalized total acoustic pressure fields $\frac{\Re{(p)}}{\max{\Re{(p)}}}$ at different angles of incidence ($\theta = 20^\circ$ and $\theta = 70^\circ$) are reported as insets.
(c, d) The maps of the difference between the absorption coefficient $\alpha_3$ of the 3$^{\rm{rd}}$ HL and $\alpha_0$ of the 0$^{\rm{th}}$ HL in the $(f,\theta)$ space for $\sigma$ = 49000 Pa.s/m$^2$ and 10000 Pa.s/m$^2$, respectively.}
\raggedright
\label{fig6}
\end{figure*}

The dependence of $\alpha$ on the angle of incidence is investigated and reported at $f$ = 500 Hz in Fig.~\ref{fig6}(a) and Fig.~\ref{fig6}(b) for $\sigma$ = 49000 Pa.s/m$^2$ (a) and $\sigma = 10000$ Pa.s/m$^2$, respectively.
Again, the introduction of hierarchy clearly shows a higher absorption coefficient in most of the angles of incidence (for both large and small values of $\sigma$), exhibiting a peak of absorption around 73$^\circ$.
The pressure fields $\frac{\Re{(p)}}{\max{\Re{(p)}}}$ for $\theta = 20^{\circ}$ and $\theta=70^{\circ}$ are reported in the insets of Fig.~\ref{fig6}(a) and Fig.~\ref{fig6}(b).
Their behaviour is very similar to those of the HP counterpart.
\\
Figure~\ref{fig6}(c) and Fig.~\ref{fig6}(d) report the maps of the difference between the absorption coefficient $\alpha_3$ of the 3$^{\rm{rd}}$ HL and $\alpha_0$ of the 0$^{\rm{th}}$ HL in the $(f,\theta)$ space for $\sigma$ = 49000 Pa.s/m$^2$ and 10000 Pa.s/m$^2$, respectively.
Contrary to the HP case, when the HG organization is considered, a significant reduction of the low-frequency region where the hierarchical design was under-performing with respect to the bulk block is observed.

\subsubsection*{Influence of the flow resistivity}
\begin{figure*}[ht!]
\centerline{\includegraphics[width= 5 in]{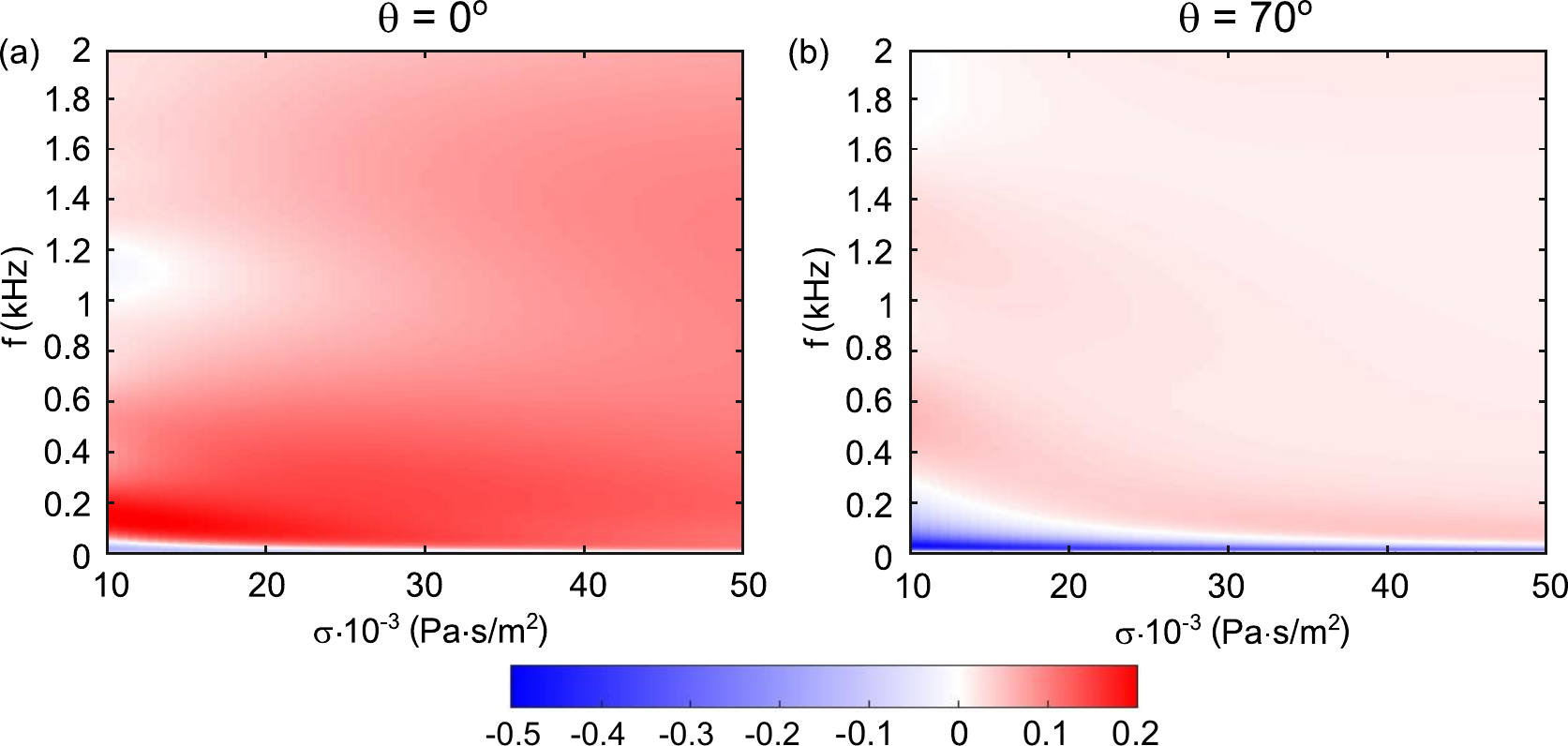}}
\caption{\textbf{Influence of the flow resistivity on the absorption coefficient}.
Difference $\alpha_3-\alpha_0$ as a function of $f$ and $\sigma$ for (a) $\theta = 0^{\circ}$ and (b) $\theta = 70^{\circ}$.}
\raggedright
\label{fig7}
\end{figure*}

Finally, the behaviour of $\alpha_3-\alpha_0$ as a function of the frequency $f$ and of the flow resistivity $\sigma$ for $\theta = 0^{\circ}$ and $\theta = 70^{\circ}$ is shown in Fig.~\ref{fig7}(a) and Fig.~\ref{fig7}(b), respectively.
Most of the $(\sigma,f)$ combinations show that the introduction of the hierarchy is beneficial in terms of absorption.
The overall increase of $\alpha$ is more consequent in the case of normal incidence (being $\alpha$ twice larger than for $\theta$ = 70$^{\circ}$ at the peak values).

\section{Conclusions}
\label{Conclusions}
We have theoretically and numerically investigated the absorption coefficient of hierarchical meta-porous media made of alternating melamine and air layers.
Two designs have been considered: a hierarchical periodic and a hierarchical gradient.
Both structures exhibit advantageous absorbing properties in the [20, 2000] Hz frequency range compared to a block of porous material of the same length.
A maximum enhancement of 39\% was observed in the hierarchical periodic organization.
Remarkably, the hierarchical geometries, optimized for a single value of flow resistivity and for normal wave incidence, only, remain advantageous designs exhibiting higher absorption in a wide range of $\theta$ (angle of incidence of the plane wave) and $\sigma$ (air flow resistivity).
Specifically, when large values of $\sigma$ are considered, a gradual increase of the absorption coefficient as additional hierarchical levels are added, has been observed for both the periodic and gradient designs.
The performance enhancement is still kept when smaller values of $\sigma$ are concerned, observing an oscillating increase of the absorption coefficient with higher peak values.
The largest enhancement of the absorption coefficient is observed at normal incidence.
The proposed hierarchical meta-porous revealed to be a promising design for increasing the low-frequency sound absorption in porous media.


\noindent\textbf{Acknowledgements}.
\noindent S.K., B.D. and M.M. acknowledge the EU, H2020 FET Open $\ll$BOHEME: Bio-Inspired Hierarchical Metamaterials$\gg$ (grant number 863179).

\vskip2pt
\bibliographystyle{RS} 
\bibliography{biblio}
\end{document}